\documentclass{iopart}
\usepackage{epsf}

\jl{6}        
\eqnobysec    

\input prepictex \input pictex \input postpictex
\def\mathput#1{\relax \ifmmode \displaystyle #1\else $\displaystyle #1$\fi}

\def\beq{\begin{equation}}
\def\eeq{\end{equation}}

\def\dualp#1{{}^{\ast_{(\hbox{$\scriptstyle #1$})}} \kern-1pt}

\def\oversymbol#1#2{\vbox{\ialign{##\crcr \hfil$#1$\hfil\crcr
   \noalign{\kern1pt\nointerlineskip}%
   \hbox{$\hfil\displaystyle#2\hfil$}\crcr}}}
\def\overcirc#1{\oversymbol{\scriptstyle\kern.5pt \circ}{#1}}
\def\four{{}^{(4)}\kern-1pt}

\begin{document}

\title[Circular Holonomy and Clock Effects in Stationary Axisymmetric Spacetimes]
{Circular Holonomy and Clock Effects in Stationary Axisymmetric Spacetimes}

\author{
Donato Bini${}^{\dag\,\ddag}$,
Robert T. Jantzen${}^{\S \,\ddag}$,
Bahram Mashhoon${}^{\P }$
}
\address{
  ${}^{\dag}$\
  Istituto per Applicazioni della Matematica C.N.R.,
  I--80131 Napoli, Italy
}
\address{
  ${}^{\ddag}$\
  International Center for Relativistic Astrophysics,
  University of Rome, I--00185 Roma, Italy
}
\address{
  ${}^{\S}$\
  Department of Mathematical Sciences, Villanova University,
  Villanova, PA 19085, USA
}
\address{
  ${}^{\P}$\
  Department of Physics and Astronomy,
University of Missouri-Columbia, Columbia
MO 65211, USA 
}

\begin{abstract}
Stationary axisymmetric spacetimes containing a pair of oppositely-rotating periodically-intersecting circular geodesics allow the study of various so-called `clock effects' by comparing either observer or geodesic proper time periods of orbital circuits defined by the observer or the geodesic crossing points. This can be extended from a comparison of clocks to a comparison of parallel transported vectors, leading to the study of special elements of the spacetime holonomy group. 
The band of holonomy invariance found for a dense subset of special geodesic orbits outside a certain radius in the static case does not exist in the nonstatic case.
In the Kerr spacetime the dimensionless frequencies  associated with parallel transport rotations can be expressed as ratios of the proper and average coordinate periods of the circular geodesics. 
\end{abstract}

\pacno{04.20.Cv}
\submitted June 28, 2001 \quad Revised October 13, 2001

\section{Introduction}

Following up the investigation of holonomy in the Schwarzschild spacetime by Rothman et al \cite{rot}, Maartens et al \cite{MMM} have  extended 
the work in \cite{rot} on closed circular trajectories
to the Kerr spacetime where the spacetime rotation further complicates matters by introducing new features.
Maartens et al \cite{MMM} study the parallel transport of a generic vector $X$ ($u$ in their notation) in the equatorial plane of the Kerr spacetime
along the orbit of the spacelike Killing vector associated with the axial symmetry.
Assuming the standard Boyer-Lindquist form of the Kerr metric with $(x^0,x^1,x^2,x^3)=(t,r,\theta,\phi)$, 
the time coordinate $t$ parametrizes each orbit of the stationary symmetry Killing vector field $\partial_t$ and
the azimuthal coordinate $\phi$ parametrizes each orbit of the axial Killing vector $\partial_\phi$.  The transport equations reduce to a single fundamental second-order ordinary differential equation for the $\phi$-derivative of the $t$-component of $X$
\beq\label{eq:MMM}
\fl\qquad
\frac{\rmd^2 Z}{\rmd\phi^2}+f^2 Z = 0 \ ,\qquad  Z=\frac{\rmd X^t }{\rmd\phi}\ ,\quad
f^2= 1-\frac{2\mathcal M}{r}-\frac{2a^2{\mathcal M}}{r^3}-\frac{a^2{\mathcal M}^2}{r^4}\ .
\label{eq:fundam}
\eeq
Recalling that $r$ is a constant along these orbits, this is trivially solved for $Z$, resulting in solutions characterized by a real or purely imaginary dimensionless frequency $f$.
The other components $X^r$,  $X^\phi$ and $X^\theta$, and $X^t$ itself are then easily calculated. 

Maartens et al \cite{MMM} discuss the holonomy question by studying the behavior of these solutions, which clearly depends on the sign of the frequency-squared function $f^2 =F/r^4$, with the roots of the quartic equation $F=0$ in $r$ distinguishing oscillatory and exponential solutions of the system. They show that $F$ has exactly one positive root $r_*$
(called $r_{\rm crit}$ in the Schwarzschild discussion
of Rothman et al \cite{rot}),
but find  no obvious significance for this value. When $F>0$, which occurs for $r>r_*$, rational values of the real frequency $f$ lead to the transported vector $X$ returning to its original value after a certain number of loops.
Here we show that $f^2$ and its sign have a geometrical significance directly associated with clock effects \cite{bijamas}
and then extend the study to the closed geodesic circuits associated with these clock effects and also studied by Rothman et al \cite{rot}
for the Schwarzschild spacetime. The results hold for any stationary axisymmetric spacetime for which clock effects can be studied. This requires the existence of pairs of oppositely-rotating timelike circular geodesics which intersect periodically, forming a double helix in spacetime contained within a single $t$-$\phi$ symmetry group orbit.

For a general family of stationary axisymmetric circularly-rotating observers, with an intermediate angular velocity between the two distinct angular velocities of such a pair of timelike circular geodesics within a given circular orbit cylinder, if the two oppositely-rotating geodesics (as seen by these observers) depart from a given observer world line simultaneously, then they will in general not return to that world line simultaneously. Instead there will be a proper time difference in their arrival times as measured by the observer clock. 
This is the observer-dependent single-clock clock effect, where the observer both defines the complete closed circuit and measures the difference in arrival times for it, interpreted as the difference in the observer-measured orbital periods. 
The observer-dependent two-clock clock effect is instead the difference in proper times elapsed on the clocks carried by the geodesics themselves for the same observer-defined circuit.
Finally the observer-independent two-clock clock effect is the difference in proper times elapsed on the clocks carried by these geodesics between three successive meeting points;
this effect is equivalent to the observer-dependent two-clock clock effect for the `geodesic meeting point observers' (GMPOs) when the latter are timelike.
The trajectories with constant angular velocity which pass through every other meeting point of the pair of geodesics are the geodesic meeting point (GMP) trajectories.
When these trajectories are timelike, they define the world lines of the GMPOs \cite{idcf1,idcf2}. 

It turns out that the zero $r_*$ of the frequency-squared function $f^2$ corresponds to the radius $r_{\rm(gmp)}$ of the observer horizon for these GMPOs. Each such observer sees the pair of corotating and counter-rotating geodesics which leave it simultaneously also return simultaneously, and thus sees vanishing single-clock clock effect.
This observer family has a number of special geometrical and kinematical properties and is also useful (by definition) in the discussion of the observer-independent two-clock clock effect.
Outside the radius $r_*$, these observers are able to `resist' the clock effect by counter-rotating sufficiently fast to synchronize the geodesic arrival times, but within this radius the rotation of the spacetime itself is too great to overcome. The GMP trajectories are well-defined whether the circular geodesics are timelike, null or spacelike; it is convenient to define their direction using their unit tangent when the tangent is non-null. Only when timelike can these GMP trajectories be interpreted as the congruence of world lines of a test observer family.

The clock effect compares the spacetime arclength for different pairs of timelike  paths between two spacetime points, each pair together forming a closed loop in spacetime. One can also compare parallel transport around such closed loops, investigating special elements of the holonomy group of the spacetime. This requires first solving the equations of parallel transport for a vector along a general circular orbit with constant angular velocity, which is done here. For comparison, the simpler case of closed $\phi$-coordinate curves is studied, followed by the loops formed by the two oppositely-rotating timelike geodesics. In the first case the geometry depends crucially on the GMPOs, while in the latter case both the GMPOs and the Carter observers come into play. The geometry of circular orbits and of the circular geodesics which give rise to both of these observer families requires preliminary study. The extremely accelerated observers \cite{bijamas,sem} also appear in this discussion as the only observers which see the parallel transport geometry symmetrically for pairs of orbits with equal magnitude but oppositely-signed relative velocities like the circular geodesics themselves.

The holonomy invariance found at a dense subset of special radii and particular loop numbers for $\phi$-coordinate loops and geodesic loops in the static case \cite{rot} and for $\phi$-coordinate loops in the stationary case \cite{MMM} and referred to as a `band' structure is reduced to a set of radii at  most of measure zero for closed geodesic loops in the nonstatic case.
In the Kerr spacetime the dimensionless frequencies associated with parallel transport rotations can be expressed as ratios of the proper and average coordinate periods of the circular geodesics. 
Establishing these results is an elegant exercise in geometry. The process of discovering them, however, initially required a computer algebra system to jumpstart the investigation and to derive certain consequences which could then be confirmed by hand. The entire discussion is a further example of how various geometrically defined circularly orbiting observer families continue to play a key role in understanding the geometry of stationary axisymmetric spacetimes.

\section{Metric and circular orbits}

Consider the spacetime metric in symmetry-adapted coordinates
\beq
\rmd s^2 = \rmd s^2_{(t,\phi)} + g_{rr}\rmd r^2 + g_{\theta\theta} \rmd\theta^2 
               \ ,
\eeq
with the metric of the circular orbit cylinder written respectively in coordinate and lapse-shift notation
\begin{eqnarray}
\fl\qquad
\label{eq:metric}
       \rmd s^2_{(t,\phi)} = g_{tt} \rmd t^2 
                    + 2 g_{t\phi} \rmd t \rmd\phi
	              +  g_{\phi\phi} \rmd\phi^2 
    = -N^2 \rmd t^2 + g_{\phi\phi}(\rmd\phi + N^\phi \rmd t)^2
 \ ,
\end{eqnarray}
where in the case of Kerr spacetime, $\{t,r,\theta,\phi \}$ are the usual Boyer-Lindquist coordinates adapted to the family of static observers, whose world lines are along the time coordinate lines, and to the zero angular momentum observers (ZAMOs), with 4-velocity 
$n=N^{-1}(\partial_t-N^\phi\partial_\phi)$ and world lines orthogonal to the time coordinate hypersurfaces. Let $e_{\hat i}=g_{ii}^{-1/2}\partial_i$ be the orthonormal frame vectors along the time slices (to be referred to as the `spherical frame'), where $i=r,\theta,\phi$. $\partial_t$ and $\partial_\phi$ are Killing vector fields associated respectively with stationarity and axisymmetry;  $\partial_\phi$ has closed orbits $\phi: 0\to2\pi$. All considerations below will be limited to the region of spacetime where 
the 1-form $\rmd t$ remains timelike and this lapse-shift decomposition remains valid, namely outside the (outer) horizon 
$r=r_+$ in the Kerr spacetime, the (outer) surface where $N=(-g^{tt})^{-1/2}\to0$. Static spacetimes occur as the special case $g_{t\phi}=0= N^\phi$.
Explicit values of the various metric quantities for the Kerr spacetime are given in Table 1 of \cite{bijamas}.

The quantity $R=g_{\phi\phi}^{1/2}$ represents the circumferential radius of the $\phi$-coordinate circles, converting $\phi$ to arclength so that the total circumference of such a circle is $2\pi R$. On the other hand
\beq
 \kappa(\phi,n)^{\hat r} 
  = - g_{rr}^{-1/2} [\ln(g_{\phi\phi}^{1/2})]_{,r} 
  = -(\ln R)_{,\hat r}
  = -1/\rho 
\eeq
is the (signed) Lie relative curvature of the $\phi$-coordinate lines with respect to the ZAMOs \cite{idcf1}
and  $\rho $ is the corresponding radius of curvature, assuming that the radial curvature component is negative, which is true outside the horizon in Kerr. The ratio 
$R/\rho = R_{,\hat r}$ is 1 in the flat spacetime case and the $r$-$\phi$ surface is just a flat plane in Euclidean $\mathcal{R}^3$.
When this ratio satisfies $0<R/\rho<1$, the $r$-$\phi$ surface can still be embedded in Euclidean $\mathcal{R}^3$ \cite{idcf2,embed}. The tangent cone to a given $\phi$-coordinate circle then has a positive deficit angle
\beq
   \Delta_{\rm(def)} = 2\pi [ 1-R/\rho ]
 \leftrightarrow
   R/\rho = 1-\Delta_{\rm(def)}/(2\pi)
\eeq
which helps gives a physical interpretation to the rotation induced by parallel transport after one circuit around the $\phi$-coordinate circle
as described below. For the Kerr spacetime, one has $0< R/\rho <1$ outside the horizon, with  $R/\rho$ increasing monotonically from its limiting value 0 at the horizon where $g_{rr}^{-1}\to0$ to its asymptotic value 1 as $r\to\infty$ \cite{idcf2}.

The trajectory of a Killing vector field $\eta=\partial_t+\zeta\partial_\phi$ is a circular orbit whose tangent vector, when non-null, can be normalized
\beq
  U(\zeta)=\Gamma(\zeta)[\partial_t+\zeta \partial_\phi]
          =\gamma(\zeta)[n+\nu(\zeta) e_{\hat\phi}]
\ , 
\eeq
where $\nu(\zeta)$ is the ZAMO relative velocity in the $\phi$ direction
and the ZAMO gamma factor is
\beq
  \gamma(\zeta)=|1-\nu(\zeta)^2|^{-1/2}
    =N\Gamma(\zeta)\ ,
\eeq
while the coordinate gamma factor $\Gamma(\zeta)$ is defined by
\begin{eqnarray}\label{eq:Q1}
\fl\qquad
 -(\partial_t+\zeta \partial_\phi) \cdot
                (\partial_t+\zeta \partial_\phi)
&=& -g_{\phi\phi} \zeta^2 
- 2g_{\phi t} \zeta 
-  g_{tt}
= -g_{\phi\phi}( \zeta-\zeta_-)(\zeta-\zeta_+) 
\nonumber\\ \fl\qquad
&\equiv& \epsilon(\zeta)^2\Gamma(\zeta)^{-2}\ , 
\end{eqnarray}
and  $\epsilon(\zeta) = [-U(\zeta) \cdot U(\zeta)]^{1/2}$ 
is respectively $1,0,i$ for timelike, null and spacelike orbits. 
The zeros $\zeta_\pm$ of this function are the angular velocities of the null circular orbits; when $\zeta=\zeta_\pm$, then $\epsilon(\zeta)=0$ and the rescaling factor $\Gamma(\zeta_\pm)$ is arbitrary.
When $\zeta_-<\zeta<\zeta_+$ so that $\epsilon(\zeta)=1$,
the quantity $\Gamma(\zeta)^{-1}$ converts intervals of coordinate time into proper time elapsed along these timelike trajectories
\beq
  \Delta \tau 
   = \Gamma(\zeta)^{-1} \Delta t 
   =  \gamma(\zeta)^{-1} N\Delta t
\ . 
\eeq

In the non-null case one can also introduce the orthogonal unit vector with angular velocity $\bar\zeta$, defining the bar map 
\begin{eqnarray}\label{eq:barnu}
  \bar U(\zeta) &=& U(\bar\zeta)
          =\gamma(\zeta)[\nu(\zeta)n + e_{\hat\phi}]
\ ,\quad
  \bar U(\zeta) \cdot \bar U(\zeta)= \epsilon(\zeta)^2\ ,
\nonumber\\
\bar\nu(\zeta) &=& \nu(\bar\zeta) = 1/\nu(\zeta)\ ,
\end{eqnarray}
which is the reflection across the null direction with the same sense of rotation as $U(\zeta)$ with respect to the ZAMOs.  
The corresponding one-form is \begin{eqnarray}\label{eq:barUY}
\bar U(\zeta)^\flat
   &=&  \gamma(\zeta)
        Y(\zeta)^\flat\ ,\
 Y(\zeta)^\flat = R (\rmd\phi -\zeta \rmd t) 
\ .
\end{eqnarray}
where $\flat$ indicates index lowering.
When $U(\zeta)$ is timelike, $\bar U(\zeta)$ is the unit vector in the $\phi$-direction in its local rest space and $\{e_{\hat r},e_{\hat\theta},\bar U(\zeta)\}$ is the boosted spherical frame. 

The relationship between the angular velocity $\zeta$ characterizing a circular orbit and the ZAMO relative velocity $\nu(\zeta)$ is
\beq
  \nu(\zeta)= N^{-1} R ( \zeta -\zeta_{\rm(nmp)} ) 
\quad\to\quad \nu_{\rm(nmp)}=\nu(\zeta_{\rm(nmp)})=0
\ ,
\eeq
where 
\beq
\zeta_{\rm(nmp)}= {\textstyle\frac12}(\zeta_+ + \zeta_-)
 =-g_{\phi t}/g_{\phi\phi} = -N^\phi
\eeq
is the angular velocity of the ZAMOs (``null meeting point observers"), whose world lines contain the meeting points of the oppositely rotating null orbits.

\section{Circular geodesics}

The circular geodesic 
4-velocities
\beq
 U_\pm = \Gamma_\pm(\partial_t +\zeta_{\rm(geo)\pm} \partial_\phi)
       = \gamma_\pm(n +\nu_\pm  e_{\hat\phi})
\eeq
are parametrized by the
angular velocities $\zeta_{\rm(geo)\pm}=U_\pm^\phi/U_\pm^t$, which
are determined by the radial component of the geodesic equation
\beq
  \frac{d U_\pm^\gamma}{d\tau} 
   + \Gamma^\gamma{}_{\alpha\beta}U^\alpha_\pm U^\beta_\pm= 0\ .
\eeq
Since $U^\alpha_\pm =dx^\alpha/d\tau$, the radial component is
\beq
  \frac{d^2 r}{d\tau^2} 
   + \Gamma^r{}_{\alpha\beta}U^\alpha_\pm U^\beta_\pm= 0\ ,
\eeq
and setting $r$ equal to a constant gives
a quadratic condition for the angular velocity
\beq\label{eq:geos}
   \Gamma^r{}_{\phi\phi} \zeta_{\rm(geo)\pm}^2 
+ 2\Gamma^r{}_{\phi t} \zeta_{\rm(geo)\pm} 
+  \Gamma^r{}_{tt}=0
\eeq
equivalent to
\beq\label{eq:Q2}
   g_{\phi\phi,r} \zeta_{\rm(geo)\pm}^2 
+ 2g_{\phi t,r} \zeta_{\rm(geo)\pm} 
+  g_{tt,r}=0\ .
\eeq
It is assumed that these two roots are distinct and satisfy $\zeta_{\rm(geo)-}\leq 0\leq \zeta_{\rm(geo)+}$, which is the case in the G\"odel spacetime in its usual stationary axisymmetric form but not in Minkowski spacetime expressed in rotating coordinates, for example. 

In order to have circular geodesics at all, it must be true that the second nonignorable coordinate can also be constant along them, leading to a condition on $\theta$ once it is set equal to a constant $\theta_{(0)}$ in the corresponding component of the geodesic equation
\beq
  \frac{\rmd^2 \theta}{\rmd\tau^2} 
   + \Gamma^\theta{}_{\alpha\beta}U^\alpha_\pm U^\beta_\pm= 0\ .
\eeq
This condition
\beq
   g_{\phi\phi,\theta} \zeta_{\rm(geo)\pm}^2 
+ 2g_{\phi t,\theta} \zeta_{\rm(geo)\pm} 
+  g_{tt,\theta}=0
\eeq
requires that these $\theta$ derivatives of these three metric components vanish at some value of $\theta$.
This is true for the equatorial plane $\theta=\pi/2$ in the Kerr spacetime because of its reflection symmetry $\theta \to \pi/2-\theta$, or for any constant value of $\theta$ in a cylindrically symmetric spacetime if $\partial_\theta$ is chosen to be the Killing vector associated with the additional translational symmetry, like the G\"odel spacetime, for example. It is also true for the
equatorial plane of {\it any} stationary axisymmetric spacetime that satisfies the
condition of reflection symmetry about the equatorial plane, i.e. 
$g_{\mu\nu}(t, r , \theta, \phi) = g_{\mu\nu} ( t, r, \pi - \theta, \phi )$; in fact, the
partial dervitative of this condition with respect to $\theta$ implies that
$g_{\mu\nu,\theta} = 0$ at $\theta = \pi/2$.
It is not true in the equatorial plane of the Kerr-Taub-NUT spacetime
\cite{mil} where circular geodesics do not exist. The remarks in \cite{bijamas} about this spacetime ignored the equations of motion for $\theta$ and are therefore incorrect when the NUT parameter $\ell$ is nonzero.

The coordinate periods 
$T_\pm =\pm 2\pi/\zeta_{\rm(geo)\pm}=2\pi/|\zeta_{\rm(geo)\pm}|$ of the two circular geodesics to complete one circuit with respect to the same static observer time line (with corresponding proper periods $\tau_\pm=\gamma_\pm^{-1} N T_\pm$ as measured by the geodesics themselves)
can be averaged to define an average coordinate period and its corresponding angular velocity
\beq \label{eq:keplerian}
\fl\qquad
T_{(0)} 
  = \pi (\zeta_{\rm(geo)_+}^{-1} - \zeta_{\rm(geo)_-}^{-1} )\ ,\
\omega_{(0)}   = 2\pi/T_{(0)}
  = 2/(\zeta_{\rm(geo)_+}^{-1} - \zeta_{\rm(geo)_-}^{-1} )
\ .
\eeq
while an asymmetry parameter can be introduced by
\beq
  a = \frac12( \zeta_{\rm(geo)_+}^{-1} + \zeta_{\rm(geo)_-}^{-1} )\ ,
\eeq
enabling one to express the $\zeta_{\rm(geo)_\pm}$ in the form
\beq 
  \zeta_{\rm(geo)\pm} = 1/(a\pm\omega_{(0)}^{-1})\ .
\eeq

In the Kerr spacetime $T_{(0)}$ and $\omega_{(0)}$ reduce to the Keplerian period and angular velocity, while the asymmetry parameter $a$ reduces to the usual rotation parameter $a$; moreover   $\omega_{(0)} = (\mathcal{M}/r^3)^{1/2}$, where $\mathcal{M}$ is the usual mass parameter and $|\omega_{(0)}|=|\tau_1|$  is the only nonvanishing Frenet-Serret torsion of these geodesics \cite{vis}.
The Kerr ZAMO relative velocities and gamma factors are explicitly
\beq\label{eq:gammaS}\fl\qquad
  \nu_\pm  =  \frac{(1\mp2a\omega_{(0)}) r^2 +a^2}{\Delta^{1/2}(a\omega_{(0)}\pm1)} \omega_{(0)} \ ,\quad
  \gamma_\pm =  \frac{\Delta^{1/2}(1\pm a\omega_{(0)})}{R(1-3\mathcal{M}/r\pm 2a\omega_{(0)})^{1/2}}\ ,
\eeq
where $\Delta=r^2-2\mathcal{M}r+a^2$ is a factor appearing in the lapse function $N=\Delta^{1/2}/R$ and in the curvature ratio 
\beq\label{eq:Kerr-condition0}
R/\rho=N(1-a^2\omega_{(0)}^2) 
\qquad\hbox{(for Kerr only)}
\,.
\eeq

With these definitions one can introduce naturally two angular velocity parameters and two time parameters, each of which has a physical interpretation. Table 1 summarizes these definitions. 

\typeout{*** Struts introduced to adjust spacing in Table 1}
  \def\UStrut{\vrule width 0pt height 12pt depth0pt}
  \def\DStrut{\vrule width 0pt height 0pt depth8pt}

\typeout{Table 1}
\begin{table}
\begin{flushright}
\begin{tabular}{|r|r|}\hline
\hbox  to .4\textwidth{\hfill
$\eqalign{\UStrut
  \zeta_{\rm(gmp)} &=(\zeta_{\rm(geo)+} + \zeta_{\rm(geo)-})/2 \cr
      &=-a\omega_{(0)}^2/(1-a^2\omega_{(0)}^2)\DStrut\cr}$\quad}
&
\hbox  to .4\textwidth{\hfill
$\eqalign{\UStrut
2\pi/\tilde T_{(0)} &=(\zeta_{\rm(geo)+} - \zeta_{\rm(geo)-} )/2 \cr
   &=\omega_{(0)}/(1-a^2\omega_{(0)}^2)
\DStrut\cr}$\quad}
\\
\hline
\hbox  to .4\textwidth{\hfill
$\eqalign{\UStrut
  \bar\zeta_{\rm(car)}^{-1} &=(\zeta_{\rm(geo)+}^{-1} + \zeta_{\rm(geo)-}^{-1})/2 \cr
      &=a = (T_+ - T_-)/2/(2\pi)\DStrut\cr}$\quad}
&
\hbox  to .4\textwidth{\hfill
$\eqalign{\UStrut
T_{(0)}/(2\pi)  &=(\zeta_{\rm(geo)+}^{-1} - \zeta_{\rm(geo)-}^{-1})/2 \cr
   &=\omega_{(0)}^{-1} \DStrut\cr}$\quad}
\\
\hline
\end{tabular}
\end{flushright}
\caption{Table of angular velocity and period parameters associated with the circular geodesics. The Carter observers are associated with the static observer clock effect.}
\end{table}

The average of the geodesic angular velocities defines the angular velocity of the GMPOs
\begin{eqnarray}
\fl\qquad
  \zeta_{\rm(gmp)} &=& (\zeta_{\rm(geo)+} + \zeta_{\rm(geo)-})/2 
  = -\Gamma^r{}_{\phi t}/\Gamma^r{}_{\phi\phi} 
  = -g_{t\phi,r}/g_{\phi\phi,r} \nonumber\\
\fl\qquad
  &=& - a\omega_{(0)}^2/(1-a^2\omega_{(0)}^2) <0\ ,
\qquad\hbox{(for $a>0$)}
\end{eqnarray}
corresponding to a counter-rotation (if $0<a<1$) with respect to the static Killing observers following the time coordinate lines. This angular velocity describes the GMP trajectories at all radii, regardless of the causal nature of the circular geodesics.
The corresponding ZAMO relative velocity is the average of the geodesic relative velocities
\beq
  \nu_{\rm(gmp)} = \nu(\zeta_{\rm(gmp)}) 
  = {\textstyle\frac12}(\nu_+ + \nu_-)\ .
\eeq

Eq.~(9.4) of \cite{bjdf} shows that the GMP trajectories are `Fermi-Walker relatively straight' with respect to the ZAMOs, which means that the direction of their relative velocity $\pm e_{\hat\phi}$ in the ZAMO rest space appears to remain constant along their trajectory compared to the directions of nearby observers: $P(n) \nabla_{U(\zeta_{\rm (gmp)})} e_{\hat \phi} =0$, where $P(n)$ is the projection into the ZAMO local rest space tangent to the time hypersurface.
This in turn shows that at the GMPO horizon, the intrinsic Lie relative curvature 
$|\kappa(\phi,n)^{\hat r}|$ of their circular orbit coincides with the single nonvanishing component
$|K(n)^{\hat r}{}_{\hat \phi}|$ of the extrinsic curvature of the time slice (the sign-reversed expansion tensor of the ZAMOs) along the orbit direction.

By definition the GMP trajectories are respectively timelike, null or spacelike when the sign $\epsilon(\zeta_{\rm(gmp)})^2$ of (\ref{eq:Q1}) takes the values $1,0,-1$. Let $r_{\rm(gmp)}$ mark the radius at which they are null: the GMPO horizon. 
In the Kerr spacetime assuming $a>0$, since $\zeta_{\rm(gmp)}<0$, these trajectories must become null before the time coordinate lines do as one approaches the horizon in the equatorial plane, so their observer horizon must occur outside the ergosphere (the time lines become null at the outer ergosphere radius).

Letting a tilde denote quantities defined for a complete geodesic circuit rather than a static observer circuit, since
\begin{eqnarray}
\fl\quad
  \tilde\zeta_{\rm(geo)\pm} 
  &=& \zeta_{\rm(geo)\pm}-\zeta_{\rm(gmp)}
  =  \pm \frac12(  \zeta_{\rm(geo)+}-\zeta_{\rm(geo)-})
= \pm\omega_{(0)}/(1-a^2\omega_{(0)}^2)
\ ,
\end{eqnarray}
the two geodesics have equal but opposite angular velocities $\tilde\zeta_{\rm(geo)+}=- \tilde\zeta_{\rm(geo)-}$ 
with respect to the GMP trajectories. Thus if two distinct circular geodesics ($\zeta_{\rm(geo)+}\neq \zeta_{\rm(geo)-})$ start at the same point on such a trajectory, they will return to it simultaneously, justifying the geodesic meeting point terminology. One such loop requires a coordinate time 
\beq\label{eq:T}
  \tilde T_{(0)}
 = 2\pi/[\frac12(  \zeta_{\rm(geo)+}-\zeta_{\rm(geo)-})]
 = T_{(0)} (1-a^2\omega_{(0)}^2)\ ,
\eeq
which corresponds to a proper time orbital period
\begin{eqnarray}\label{eq:2period}
\fl\qquad
 \tilde\tau_{(0)}
  &=& \gamma(\zeta_{\rm(gmp)})^{-1} N \tilde T_{(0)} \nonumber\\
\fl\qquad
&=& |1-2\mathcal{M}/r-a^2\omega_{(0)}^2(2+\mathcal{M}/r)|^{1/2} T_{(0)}
\qquad\hbox{(for Kerr only)}
\end{eqnarray}
measured by the GMPOs and to 
\begin{eqnarray}
  \tilde\tau_\pm
  &=& \gamma_\pm^{-1} N \tilde T_{(0)}\nonumber\\
  &=& \frac{(1-3\mathcal{M}/r\pm 2a\omega_{(0)})^{1/2}}{1\pm a\omega_{(0)}}  \tilde T_{(0)} 
\qquad\hbox{(for Kerr only)}
\end{eqnarray}
as measured by the geodesics themselves.
The difference of the latter two proper periods is the observer-independent two-clock clock effect
\beq\label{eq:2c}
\fl\qquad
  \tilde\tau_+ - \tilde\tau_- = (\gamma_+^{-1}-\gamma_-^{-1}) N \tilde T_{(0)}
= (\gamma_+^{-1}-\gamma_-^{-1}) \gamma(\zeta_{\rm(gmp)})\tilde\tau_{(0)}
   \ .
\eeq
It is worth noting the following period ratio
\begin{eqnarray}\label{eq:Kerr-condition}
    \frac{\tilde T_{(0)}}{T_{(0})}
&=&  - 4 \frac{\zeta_{\rm(geo)+} \zeta_{\rm(geo)-} }
               { (\zeta_{\rm(geo)+} - \zeta_{\rm(geo)-})^2}
 = 1-a^2\omega_{(0)}^2
         \nonumber\\  
&=&  \frac{R/\rho}{N}\ ,\qquad \hbox{(for Kerr only)}
\end{eqnarray}
where the last equality follows from  Eq.~(\ref{eq:Kerr-condition0}).

Finally the product
\beq
    \zeta_{\rm(gmp)} \tilde T_{(0)}
      = 2\pi \frac{\zeta_{\rm(geo)_+} + \zeta_{\rm(geo)_-} }
             {\zeta_{\rm(geo)_+} - \zeta_{\rm(geo)_-} } 
  = - 2\pi a \omega_{(0)}\ ,
\eeq
is the change in coordinate angle $\phi$ by which the GMPO has moved during one such circuit, itself the average of the individual angles that the geodesics have moved during this circuit
\beq
    \zeta_{\rm(geo)\pm} \tilde T_{(0)}
      = 4\pi \frac{\zeta_{\rm(geo)_\pm} }
             {\zeta_{\rm(geo)_+} - \zeta_{\rm(geo)_-} } 
  = \pm 2\pi (1 \mp a \omega_{(0)})\ ,
\eeq
so $a \omega_{(0)}$ and $1 \mp a \omega_{(0)}$ are the number of circuits with respect to the static observers made by the GMPOs and geodesics respectively corresponding to each geodesic loop. 
For the Kerr spacetime,
one may express the circuit number $a\omega_{(0)}$ of the GMPOs themselves in terms of the angular velocity 
$\Omega_{\rm(h)}= a/(r_+^2 + a^2)$ of the black hole \cite{mtw}
\beq
  a\omega_{(0)}|_{r=r_+} = \mathcal{M}\Omega_{\rm(h)}[2/(1-a\Omega_{\rm(h)})]^{1/2} \ ,
\eeq
which is a proper fraction for $0\leq a/\mathcal{M} <1$ but has the value 1 for the extreme case $a/\mathcal{M} =1$ where the corotating circular geodesic becomes null at the horizon.

Along a general  circular orbit of angular velocity $\zeta$ starting at $t=0,\phi=0$ and parametrized by $t$, one has $\phi=\zeta t$. The continuous parameter $\phi$ along the orbit is a parameter taking all real values in order to describe any number of circuits around the $\phi$-coordinate circle in either direction. Its value mod $2\pi$ gives a unique coordinate angle in the interval $[0,2\pi)$ for a point on such a curve.
If one introduces a new angular coordinate $\tilde\phi=\phi- \zeta_{\rm(gmp)}t$ 
which is comoving with respect to the GMPOs,
then this induces a new parametrization $\tilde\phi=(\zeta - \zeta_{\rm(gmp)}) t =\tilde\zeta t $, with $\tilde\phi: 0\to\pm2\pi$ describing one complete circuit corotating or counterrotating with respect to the GMPOs.
The two angular parametrizations are related to each other by 
\beq\label{eq:tildephi}
  \tilde\phi 
       = (1- \zeta_{\rm(gmp)}/\zeta) \phi\ .
\eeq
For the circular geodesics this becomes
\beq\label{eq:tildephigeo}
  \tilde\phi 
       = (1- \zeta_{\rm(gmp)}/\zeta_{\rm(geo)\pm}) \phi
       = (1\mp a\omega_{(0)}) \phi
\ .
\eeq

\typeout{Table 2}
\begin{table}
\begin{center}
\begin{tabular}{|l|l|l|l|l|l|}\hline
 & null & geos & ZAMOs & GMPOs
\\ \hline
$\zeta$ & $\zeta_\pm$ & $\zeta_{\rm(geo)\pm}$ & $\zeta_{\rm(nmp)} =\frac12(\zeta_++\zeta_-)$ & $\zeta_{\rm(gmp)} =\frac12(\zeta_{\rm(geo)+} +\zeta_{\rm(geo)-})$ 
\\
$\nu(\zeta)$ & $\pm1$ & $\nu_{\rm(geo)\pm}$ & $\nu_{\rm(nmp)} =0$ & $\nu_{\rm(gmp)} = \frac12(\nu_+ + \nu_-)$ 
\\
\hline
\end{tabular}
\end{center}
\caption{Table of angular velocities and corresponding ZAMO relative velocities.}
\end{table}

Table 2  summarizes the basic angular velocities and their corresponding ZAMO relative velocities instroduced so far.
The average of the reciprocals of the geodesic angular velocities defines the reciprocal angular velocity of the orthogonal direction 
within the $t$-$\phi$ cylinder
to another important observer family, the Carter observers
\begin{eqnarray}
\fl\qquad
  1/\bar\zeta_{\rm(car)} 
  &=& (\zeta_{\rm(geo)+}^{-1} + \zeta_{\rm(geo)-}^{-1})/2 
  = -\Gamma^r{}_{\phi t}/\Gamma^r{}_{tt} 
  = -g_{t\phi,r}/g_{tt,r} = a\ .
\end{eqnarray}
Note that in the static case $g_{t\phi}=0$, one has
$\zeta_{\rm(gmp)}=0 = \zeta_{\rm(car)}= \bar\zeta_{\rm(car)}^{-1}$.
In the Kerr spacetime, the Carter observers and the ZAMOs share the same observer horizon, which is the (outer) horizon of the black hole.

Another interesting observer family, the extremely accelerated observers \cite{bijamas,sem}, extremizes the magnitude of the acceleration among the family of circularly rotating observers at each radius.
Their angular velocity $\zeta_{\rm(ext)}$ is given by a coordinate gamma factor weighted average of the geodesic angular velocities
\beq\label{eq:zetaeao}
\zeta_{\rm(ext)} 
  =(\Gamma_+\zeta_{\rm(geo)+} + \Gamma_-\zeta_{\rm(geo)-})/(\Gamma_+ + \Gamma_-)\ .
\eeq

\section{Angular velocity maps}

The quadratic forms (\ref{eq:Q1}) and (\ref{eq:Q2}) may be associated with inner products on the space of spacetime angular velocity vectors $\partial_t +\zeta \partial_\phi$ describing Killing trajectories. In both cases the condition of orthogonality with respect to the inner product 
\beq
\fl\qquad
  \left(\begin{array}{cc} 1 & \zeta_2 
  \end{array}\right)
  \left(\begin{array}{cc}C & B \\ B & A 
  \end{array}\right)
  \left(\begin{array}{c}1\\ \zeta_1 
  \end{array}\right)
 = A \zeta_1\zeta_2 + B(\zeta_1+\zeta_2) + C
 = 0
\eeq
leads to a fractional linear transformation between the two angular velocities
\beq
  \zeta_2 = -\frac{C+B\zeta_1}{B+A\zeta_1}
\mathop{\longrightarrow}\limits^{B\to0}
 -\frac{C}{A} \frac{1}{\zeta_1}
\eeq
which is involutive: $\zeta_1 \to \zeta_2 \to \zeta_1$.

For metric orthogonality
$(A,B,C)=(g_{\phi\phi},g_{\phi t},g_{tt})$ this defines the bar map
\beq
 \bar\zeta 
    = \zeta_{\rm(nmp)} 
   \frac{\zeta-\bar\zeta_{\rm(th)}}{\zeta-\zeta_{\rm(nmp)}}
\mathop{\longrightarrow}\limits^{g_{\phi t}\to0}
 -\frac{g_{tt}}{g_{\phi\phi}} \frac{1}{\zeta}
\ ,
\eeq
giving the angular velocity of the direction orthogonal to the starting direction, where
$\bar\zeta_{\rm(th)}=-1/(g_{\phi t}/g_{tt})$ 
is the angular velocity of the direction orthogonal to the static Killing observers. The bar map was already shown to be the reciprocal map when expressed in terms of ZAMO relative velocity in (\ref{eq:barnu}). Note that the null directions are invariant under this reciprocal map, so 
$\bar\zeta_\pm = \zeta_\pm$.

For vanishing radial covariant derivative of one with respect to the other,
\beq
\fl\qquad
  -2[\nabla_{\partial_t+\zeta\partial_\phi} (\partial_t+\mathcal{Z}(\zeta)\partial_\phi)]_{r} 
 = A \zeta \, \mathcal{Z}(\zeta) + 2B[\zeta+\mathcal{Z}(\zeta)] + C =0\ ,
\eeq
which is the quadratic form corresponding to the condition (\ref{eq:Q2}) satisfied by the geodesic angular velocities,
then 
$(A,B,C)=(g_{\phi\phi,r},g_{\phi t,r},g_{tt,r})$ and this defines a new map $\mathcal{Z}$
\beq\label{eq:Zzeta}
 \mathcal{Z}(\zeta) 
    = \zeta_{\rm(gmp)} 
   \frac{\zeta-\bar\zeta_{\rm(car)}}{\zeta-\zeta_{\rm(gmp)}}
\mathop{\longrightarrow}\limits^{g_{\phi t}\to0}
 -\frac{g_{tt,r}}{g_{\phi\phi,r}} \frac{1}{\zeta}
\ ,\eeq
satisfying $\mathcal{Z}(\mathcal{Z}(\zeta))=\zeta$.
$\mathcal{Z}(\zeta)$ is the angular velocity of the stationary axisymmetric vector field tangent to the symmetry orbit which is covariant constant along the Killing trajectory with angular velocity $\zeta$.

Evaluating the map $\mathcal{Z}$ in terms of the ZAMO relative velocity,
one finds
\begin{eqnarray}\label{eq:Omeganu}
 \nu(\mathcal{Z}(\zeta)) 
   &=&\nu_{\rm(gmp)}
    \frac{\nu(\zeta)-\nu_{\mathcal{Z}\rm(nmp)}}{\nu(\zeta)-\nu_{\rm(gmp)}}
\quad 
\mathop{\longrightarrow}\limits^{g_{\phi t} \to0}
\quad
-\nu_+ \nu_-/\nu(\zeta)
\ ,\nonumber\\
 \gamma(\mathcal{Z}(\zeta)) 
&=&\frac{\gamma_{\rm(gmp)}|\nu-\nu_{\rm(gmp)}|}
      {|[\nu-\nu(\mathcal{Z}(\zeta_+))][\nu-\nu(\mathcal{Z}(\zeta_-))]|^{1/2}} \ ,
\end{eqnarray}
where the vanishing ZAMO relative velocity
$\nu(\zeta_{\rm(nmp)})=0$ of the ZAMOs themselves shows that
\beq
\nu_{\mathcal{Z}\rm(nmp)} 
= \nu(\mathcal{Z}(\zeta_{\rm(nmp)}))
= [\nu_+^{-1}+\nu_-^{-1}]^{-1}/2 \ ,
\eeq
the explicit value following from the actual calculation.

Since $\mathcal{Z}(\mathcal{Z}(\zeta))=\zeta$, the vector with angular velocity $\mathcal{Z}(\zeta)$ is null at
the angular velocities
\beq\label{eq:Zzeta2}
\mathcal{Z}(\zeta_\pm)  
= \zeta_{\rm(gmp)} \frac{\zeta_\pm-\bar\zeta_{\rm(car)}}{\zeta_\pm-\zeta_{\rm(gmp)}}
\ ,
\eeq
or equivalently in relative velocity form
\beq
\label{eq:Znull}
\nu(\mathcal{Z}(\zeta_\pm))
  =\frac{\nu_{\mathcal{Z}\rm(nmp)}\mp1}{1 \mp1/\nu_{\rm(gmp)}}
  = \frac{2\nu_+\nu_- 
           \mp (\nu_+ + \nu_-)}
         {\nu_+ + \nu_- \mp 2}
\,\mathop{\longrightarrow}\limits^{g_{\phi t} \to0}\,
 \mp \nu_+ \nu_-
\ .
\eeq
From this one easily derives the inequalities
\beq
\nu_- < \nu(\mathcal{Z}(\zeta_-))  <
 \nu(\mathcal{Z}(\zeta_+)) < \nu_+\ ,
\eeq
with equality holding for the outer inequalities in the separate limits $\nu_\pm  \to \pm 1$ when the corresponding geodesics become null. These outer inequalities separately reverse when the corresponding geodesics are spacelike. Thus in the interval
$\mathcal{Z}(\zeta_-)  < \zeta < \mathcal{Z}(\zeta_+)$ or equivalently
$\nu(\mathcal{Z}(\zeta_-))  < \nu(\zeta) < \nu(\mathcal{Z}(\zeta_+))$,
the vector $\partial_t+\mathcal{Z}(\zeta)\partial_\phi$ which is parallel transported along the curve with angular velocity $\zeta$ is spacelike, and outside this interval it is null (at the endpoints) and then timelike. The bar map sends this vector to an orthogonal one whose 1-form according to (\ref{eq:barUY}) is proportional to
\beq 
  Y(\mathcal{Z}(\zeta))^\flat = R [d\phi -\mathcal{Z}(\zeta) dt]\ ,
\eeq
which is then timelike within this interval and spacelike outside it, while remaining null at the endpoints.

\section{Closed circular orbit holonomy}

Parallel transport of a vector $X$ around the $\phi$-coordinate circles (parametrized by the coordinate $\phi$)
is governed by the constant coefficient linear system  of differential equations for the coordinate components
\beq\label{eq:A}
\nabla_\phi X^\alpha =0 \longleftrightarrow 
\frac{\rmd X^\alpha}{\rmd\phi}  -A^\alpha{}_\mu X^\mu =0\ ,
\eeq
which has the general solution
\beq
X(\phi) =e^{\phi A} X(0)\ .
\eeq
The matrix 
$A^\alpha{}_\mu = -\Gamma^\alpha{}_{\phi\mu} 
= -g^{\alpha \beta}g_{\phi [\beta ,\mu]}$, which consists of the coordinate components of a mixed tensor $A$ whose covariant components $A_{\alpha\beta} =-A_{\beta\alpha}$ are antisymmetric and therefore define a 2-form, generates a 1-parameter family of Lorentz transformations of the tangent space expressed in nonorthonormal coordinates on that space. This matrix (constant along the $\phi$-coordinate circles) evaluates explicitly to
\beq\label{eq:AY}
A^\alpha{}_\mu 
=  (\ln R)_{, r}[\rmd r\wedge Y(\zeta_{\rm(gmp)})^\flat]^\alpha{}_\mu
= R/\rho  \,
[e_{\hat r}^\flat \wedge Y(\zeta_{\rm(gmp)})^\flat]^\alpha{}_\mu\ .
\eeq

Since this is the mixed index form of a decomposable 2-form (wedge product of two 1-forms), interpretating its exponential as a Lorentz transformation is simple. When nonnull, decomposing this 2-form into its magnitude $\sigma$ and a unit 2-form $B$ (with unit forms suitably defined) yields the orientation of the 2-plane (spanned by the corresponding vectors) in which either a boost (timelike case) or rotation (spacelike case) takes place and the corresponding pseudoangle or angle is that magnitude $\sigma$, while in the null case, normalization is not possible and a null rotation takes place in this plane.
The following factorization does just that
\beq
A^\alpha{}_\mu 
= \sigma B^\alpha{}_\mu
= \sigma_{(0)} \mathcal{B}^\alpha{}_\mu
 \ ,
\eeq
where recalling the notation of (\ref{eq:barUY}) one defines
\begin{eqnarray}
\sigma 
&=& \gamma(\zeta_{\rm(gmp)})^{-1} R/\rho  
         \ ,\quad
\sigma_{(0)} = R/\rho \ ,\nonumber\\
B^\flat 
&=& e_{\hat r}^\flat \wedge \bar U(\zeta_{\rm(gmp)})^\flat \ ,\quad
\mathcal{B}^\flat =
  e_{\hat r}^\flat \wedge Y(\zeta_{\rm(gmp)})^\flat\ ,
\end{eqnarray}
with the properties
\beq\label{eq:BB}
B^3 
= -\epsilon(\zeta_{\rm(gmp)})^2 B\ ,\quad  
\mathcal{B}^3 
= -\epsilon(\zeta_{\rm(gmp)})^2 \gamma(\zeta_{\rm(gmp)})^{-2} \mathcal{B}\ .
\eeq
Recall that the unit vector $\bar U(\zeta_{\rm(gmp)})$ when spacelike is just the $\phi$ direction in the GMPO local rest space.

Thus the 2-plane of this Lorentz transformation is spanned by the radial direction and the direction within the $t$-$\phi$ cylinder orthogonal to the GMP trajectory direction, with the factor $\epsilon(\zeta_{\rm(gmp)})^2 = 1,0,-1$ describing these three cases in which the 2-plane is spacelike, null, and timelike respectively, corresponding to a rotation, null rotation, and boost. Thus outside the GMP horizon in a Kerr spacetime, this is a rotation in the $r$-$\phi$ plane in the GMPO local rest space and the identity in the orthogonal 2-plane spanned by $U(\zeta_{\rm(gmp)})$ and $e_{\hat\theta}$.

The exponentiation is easily evaluated by collapsing the infinite series using the properties (\ref{eq:BB}); letting 1 stand for the identity matrix in the exponentiation, one finds
\beq\label{eq:expB}
\fl\qquad
 e^{\phi A} 
 = \cases{ 
e^{\phi\sigma B}
= [1-\mathcal{P}]
  + \cos[\epsilon(\zeta_{\rm(gmp)})\sigma\phi] 
  \, \mathcal{P} &\cr
  \hskip30pt + \epsilon(\zeta_{\rm(gmp)})^{-1} 
    \sin[\epsilon(\zeta_{\rm(gmp)})\sigma\phi] 
  \,  B \ , 
   & \hbox{$\epsilon(\zeta_{\rm(gmp)})\neq0$\,,}\cr
e^{\phi\sigma_{(0)} \mathcal{B}}
=1 + \sigma_{(0)}\phi \mathcal{B} + \frac12\sigma_{(0)}^2\phi^2 \mathcal{B}^2\ , 
   & 
\hbox{$\epsilon(\zeta_{\rm(gmp)})=0$\,,}\cr
}
\eeq
where the matrix / mixed tensor
\beq
 \mathcal{P} = -\epsilon(\zeta_{\rm(gmp)})^{-2}B^2 \ ,\quad 
\epsilon(\zeta_{\rm(gmp)})\neq0\ ,
\eeq
satisfying $\mathcal{P}^2=\mathcal{P}$ and $B\mathcal{P}=B=\mathcal{P}B$,
is the projection into the 2-plane spanned by $e_{\hat r}$ and $\bar U(\zeta_{\rm(gmp)})$.
The exponential matrix represents a Lorentz transformation acting in this 2-plane in the nonnull case, while acting as the identity in the orthogonal 2-plane with projector $1-\mathcal{P}$.
When
$\epsilon(\zeta_{\rm(gmp)})=1$, it is a clockwise (negative) rotation by the angle $\sigma\phi$ with respect to the ordered orthonormal axes $(e_{\hat r},\bar U(\zeta_{\rm(gmp)}))$ of the boosted spherical frame.
Note that $\partial_t+\zeta_{\rm(gmp)}\partial_\phi$ is the direction of the 1-parameter family of stationary axisymmetric vectors tangent to the $t$-$\phi$ cylinder which is invariant under this transformation and hence is covariant constant along the curve.

The eigenvalues of the matrix $A$ are $0,0,\pm f$, where the frequency function $f=\epsilon(\zeta_{\rm(gmp)})\sigma$ can be real, zero or purely imaginary. One may also obtain decoupled equations by differentiation and backsubstitution instead of the eigenvalue-eigenvector approach (equivalent to the above explicit exponentiation). Thus
\beq\label{decoupled}
   \frac{\rmd^2}{\rmd\phi^2}\frac{\rmd X^\alpha}{\rmd\phi} 
 = \sigma^3 [B^3 X]^\alpha
 = -f^2 [\sigma B X]^\alpha
 = -f^{2} \frac{\rmd X^\alpha}{\rmd\phi}
\eeq
leads to decoupled equations for each of the derivatives $dX^\alpha/d\phi$, for example. For the Kerr spacetime, setting $\alpha=t$ leads to Eq.~(\ref{eq:MMM}) used by Maartens et al \cite{MMM}.

Note that for asymptotically flat spacetimes like Kerr, as $r\to\infty$ one has the limits $(g_{rr},g_{\phi\phi})\to (1,r^2)$, $\rho \to r$, $\gamma(\zeta_{\rm(gmp)})\to 1$, and $\sigma\to 1$ leading to an angular velocity of $1$ in the $r$-$\phi$ plane of the tangent space, while $B$ generates a clockwise rotation of the ordered boosted $r$ and $\phi$ axes. This causes a parallel transported vector to rotate in this plane by an angle $\phi$ in the clockwise direction with respect to the boosted coordinate axes to maintain its direction as those axes rotate in the counterclockwise direction by the same angle as $\phi$ increases. One must subtract this orbital rotation by the angle $\phi$ in order to obtain the rotation of the parallel transported axes with respect to the nonrotating axes at spatial infinity, corresponding to the angular velocity $\sigma-1$, or $1-\sigma$ in the usual counterclockwise direction.

The dimensionless frequency function can be expressed in the form (see Eq.~(\ref{eq:Kerr-condition}))
\begin{eqnarray}\label{eq:Tratio}
\fl\qquad
  \sigma&=& \tilde\tau_{(0)} \frac{R/\rho}{N \tilde T_{(0)}}
      = \frac{\tilde\tau_{(0)}}{T_{(0)}} \left( \frac{R/\rho}{N} \frac{T_{(0)}}{\tilde T_{(0)}} \right)
\nonumber\\
\fl\qquad
 &=&\frac{\tilde\tau_{(0)}}{T_{(0)}} 
  = \left[1-\frac{2\mathcal{M}}{r} -a^2\omega_{(0)}^2 (2+\frac{\mathcal{M}}{r})\right]^{1/2}
\qquad\hbox{(for Kerr only)}\,.
\end{eqnarray}
It converts the angle of revolution of the geodesic with respect to the static observers to the corresponding angle that a parallel transported vector rotates in the opposite direction with respect to the spherical frame about a fixed axis in the GMPO local rest space, and in the Kerr spacetime it is just the ratio of the GMP proper geodesic period to the average static observer coordinate period for the geodesic pair.

In the combination
\beq
  \sigma\phi 
=  \gamma(\zeta_{\rm(gmp)})^{-1} [R \phi]/
        \rho \ ,
\eeq
representing the total parallel transport angle from 0 to $\phi$ (relative to the spherical frame),
the bracketed expression is the arclength along the coordinate circle and the gamma factor converts this to the arclength along the corresponding circular trajectory orthogonal to the GMP trajectories (Lorentz contraction when timelike) which projects down to the coordinate circle along the ZAMO trajectories, and finally  the last factor takes its ratio with respect to the local radius of curvature of the coordinate circle as seen by the ZAMOs. The angular velocity goes to zero exactly where the gamma factor goes to infinity at $r=r_{\rm(gmp)}$, the horizon for the GMPOs.
This relation can be restated in terms of the embedding deficit angle as
\beq
\sigma
= \gamma(\zeta_{\rm (gmp)})^{-1} (1-\frac{\Delta_{\rm(def)}}{2\pi}) \ .
\eeq
Then for a change of angle $\phi=\pm2\pi$
corresponding to one revolution, one has
\begin{eqnarray}
 \sigma\phi &=& \pm\gamma(\zeta_{\rm (gmp)})^{-1}(2\pi-\Delta_{\rm(def)})\ ,
\nonumber\\
 (\sigma-1)\phi &=&  \pm2\pi[\gamma(\zeta_{\rm (gmp)})^{-1}-1]
 \mp\gamma(\zeta_{\rm (gmp)})^{-1}\Delta_{\rm(def)}\ .
\end{eqnarray}
Since $B$ generates a clockwise (negative) rotation with respect to the ordered axes
$(e_{\hat r},\bar U(\zeta_{\rm(gmp)}))$ of the boosted spherical frame, the corotating/counterrotating circuit leads to a counterclockwise/clockwise rotation of a parallel transported vector with respect to these two axes by the deficit angle $\Delta_{\rm(def)}$ when the gamma factor is close to 1, corresponding to Figure 2 of Thorne \cite{thorne}.

Where both circular geodesics are timelike and $\sigma$ is positive
and has a rational value $\sigma=p/q$ at a given radius,
where $p$ and $q$ are positive integers, then after $q$ revolutions ($\phi=\pm 2\pi q$), any vector undergoes a rotation by $\sigma\phi=\pm 2\pi p$ and will thus
rotate back to its original starting value (`holonomy invariance') so the holonomy group restricted to this family of curves (fixed $p$, variable $q$)
must reduce to a discrete subgroup of the rotation group. For the Kerr spacetime $\sigma$ is a monotonically increasing function of $r$ from its zero value at the GMPO horizon $r_{\rm(gmp)}$ to its asymptotic limiting value 1 at $r\to\infty$, so if $\sigma=p/q$, then $p<q$ and $q>1$. This means that a transported vector undergoes fewer revolutions with respect to the boosted spherical frame than the number of loops around the hole before returning to its original orientation. The radii at which such holonomy invariance occurs for the `harmonic' parameter family $q=2,3,\ldots$, $p=1,2,\ldots,q-1$ cover the interval $(r_{\rm(gmp)},\infty)$ densely in the same way that the rational proper fractions cover the interval $(0,1)$, creating the band structure referred to in \cite{rot}.

\section{General circular orbit transport}

In order to compare the parallel transport of a vector after it returns to the same GMP trajectory after being carried along by the two oppositely-rotating circular geodesics, one first needs to extend this discussion to circular orbits with constant angular velocity $\zeta\in R$. 
These curves can be parametrized naturally by either the time coordinate
$t$ or when $\zeta\neq0$ by the extended values $\phi=\zeta t$ of the $\phi$-coordinate, assuming such a parametrized curve starts at $t=0$ and $\phi=0$ for simplicity. Corotating and counterrotating orbits (of any number of circuits with respect to the reference orbits of the static observers with $\zeta=0$ following the time coordinate lines) correspond to positive and negative angular velocities respectively. 
Note that the previous case of $\phi$-coordinate circles corresponds to the limits
$ \lim_{\zeta\to\pm\infty} \zeta^{-1} (\partial_t+\zeta\partial_\phi) 
  = \partial_\phi$.
One can also use a continuous parametrization corresponding to the comoving angular coordinate $\tilde\phi=(1- \zeta_{\rm(gmp)}/\zeta)\phi$ relative to the GMPOs as a parameter along this orbit.

The equation $\nabla_{U(\zeta)} X =0$ of parallel transport along such a parametrized curve has the same form as  (\ref{eq:A})
\beq\label{eq:Azeta}
\fl\qquad
(\nabla_t+\zeta\nabla_\phi) X^\alpha =0 \longleftrightarrow
\cases{
\frac{\rmd X^\alpha}{\rmd\phi}  -A(\zeta)^\alpha{}_\mu X^\mu =0\ , 
&$\zeta\neq0$\ ,\cr
\frac{\rmd X^\alpha}{\rmd t}  -\zeta A(\zeta)^\alpha{}_\mu X^\mu =0\ ,
& $\zeta=0$\ ,}
\eeq
where the matrix 
\beq
  A(\zeta)^\alpha{}_\mu
   = -(\Gamma^\alpha{}_{\phi\mu}+ \zeta^{-1}\Gamma^\alpha{}_{t\mu}) 
   = -\zeta^{-1} g^{\alpha \beta}(g_{t [\beta ,\mu]}+\zeta g_{\phi [\beta ,\mu]})
\eeq
is again antisymmetric when index-lowered and so corresponds to
the mixed tensor form of a 2-form and generates a Lorentz transformation, although expressed in nonorthonormal coordinates on the tangent space.
It is obtained from $A$ in the new derivation adding an extra term to $Y(\zeta_{\rm(gmp)})^\flat=R[\rmd\phi-\zeta_{\rm(gmp)} \rmd t]$ in (\ref{eq:AY}) yielding instead
\begin{eqnarray}
&& 
  R[\rmd\phi-\zeta_{\rm(gmp)} \rmd t 
         -(\zeta_{\rm(gmp)}/\zeta)(\rmd\phi-\bar\zeta_{\rm(car)} \rmd t)]
\nonumber\\ 
\qquad
&& =\cases{
(1-\zeta_{\rm(gmp)}/\zeta) Y(\mathcal{Z}(\zeta))^\flat\ , 
        & $\zeta\neq \zeta_{\rm(gmp)}$\ ,\cr
 (\bar\zeta_{\rm(car)} - \zeta_{\rm(gmp)}) R \rmd t\ , 
        & $\zeta = \zeta_{\rm(gmp)}$\ ,\cr 
 ( 1-\zeta_{\rm(gmp)}/\bar\zeta_{\rm(car)}) R \rmd\phi\ , 
        & $\zeta = \bar\zeta_{\rm(car)}$\ ,\cr
}
\end{eqnarray}
thus interpolating between  $d\phi$ (when $\zeta=\bar\zeta_{\rm(car)}$) and $dt$ (when $\zeta=\zeta_{\rm(gmp)}$).

The angular velocity map $\mathcal{Z}$ of (\ref{eq:Zzeta})
determines the trajectories orthogonal to the plane of the Lorentz transformation. Direct calculation shows that it
satisfies
\begin{eqnarray}\label{eq:Zconditions}
\fl\quad
 &&
\mathcal{Z}(\zeta_{\rm(geo)\pm}) = \zeta_{\rm(geo)\pm}\ ,\
\mathcal{Z}(\zeta_{\rm(ext)}) = \bar\zeta_{\rm(ext)}\ ,\
\mathcal{Z}(\bar\zeta_{\rm(ext)}) = \zeta_{\rm(ext)}\ , 
\nonumber\\
\fl\quad
 &&\mathcal{Z}(0) = \bar\zeta_{\rm(car)}\ ,\
  \mathcal{Z}(\bar\zeta_{\rm(car)})=0=  \mathcal{Z}(\zeta_{\rm(gmp)})^{-1} 
\ ,\ \lim_{\zeta\to\pm\infty}\mathcal{Z}(\zeta) = \zeta_{\rm(gmp)}
 \ ,
\end{eqnarray}
where $\zeta_{\rm(ext)}$ is the angular velocity of the extremely accelerated observers (\ref{eq:zetaeao}).
The limits $\zeta\to\pm\infty$ reduce the present case to the previous case for the $\phi$-coordinate circles, while for $\zeta=0$ this shows that along the time coordinate lines, the Carter observer local angular trajectories play the same role that the GMP trajectories play for the $\phi$-coordinate circles.
The boost from $U(\zeta)$ to $U(\mathcal{Z}(\zeta))$ when both are timelike maps the local rest space of the orbit along $U(\zeta)$ to the local rest space of the plane in which the parallel transport Lorentz transformation takes place. The vector $\bar U(\mathcal{Z}(\zeta))$ is a unit vector along the $\phi$-direction in this latter local rest space and $\{e_{\hat r},e_{\hat\theta},\bar U(\mathcal{Z}(\zeta))\}$ is the boosted spherical frame.
Figure 1 shows the orientation of the intersection of the Lorentz transformation plane with the $t$-$\phi$ tangent plane 
(i.e.\ $\bar U(\mathcal{Z}(\zeta))=U(\bar\mathcal{Z}(\zeta))$ modulo sign)
for the directions of the various 4-velocities which arise in this discussion.

\begin{figure}[t]
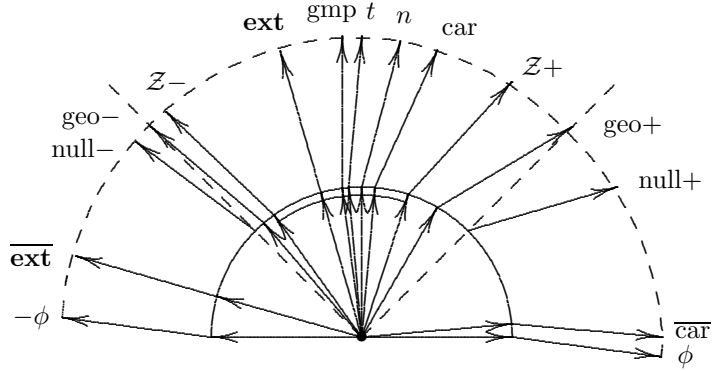

\typeout{Figure 1}
$$ \vbox{
\beginpicture
\setcoordinatesystem units <2cm,2cm> point at  0 0  

\circulararc 180 degrees from 1 0 center at 0 0

\put {\mathput{\bullet}}                at  0 0

\circulararc 54.5 degrees from .29 .90 center at 0 0

\setdashes 
\plot 0 0 1.71 1.71 /
\plot 0 0 -1.71 1.71 / 
\setsolid 

\arrow <.35cm> [.1,.4] from  0 0 to 0.50 0.86  
\arrow <.35cm> [.1,.4] from  .50 .86 to 1.4 1.4
\arrow <.35cm> [.1,.4] from  0 0 to  -.65 .76  
\arrow <.35cm> [.1,.4] from  -.65 .76 to -1.4 1.4
\arrow <.35cm> [.1,.4] from  0 0 to  -.13 .99  
\arrow <.35cm> [.1,.4] from  -.13 .99 to -.13 2.0 
\arrow <.35cm> [.1,.4] from  0 0 to  .088  .99  
\arrow <.35cm> [.1,.4] from .088  1.0 to .503 1.91
\arrow <.35cm> [.1,.4] from  0 0 to  1.0 .088  
\arrow <.35cm> [.1,.4] from  1.0 .088 to 2.0 .0013 
\arrow <.35cm> [.1,.4] from  0 0 to  -.27 .96  
\arrow <.35cm> [.1,.4] from  -.27 .96  to -.54 1.9  
\arrow <.35cm> [.1,.4] from  0 0 to  -.96 .27  
\arrow <.35cm> [.1,.4] from  -.96 .27  to -1.9 .54    
\arrow <.35cm> [.1,.4] from  .71 .71 to 1.7 1.0 
\arrow <.35cm> [.1,.4] from  -.71 .71 to -1.5 1.3
\arrow <.35cm> [.1,.4] from  0 0 to  .31  .95 
\arrow <.35cm> [.1,.4] from  .31  .95 to 1.0 1.7
\arrow <.35cm> [.1,.4] from  0 0 to  -.58 .81  
\arrow <.35cm> [.1,.4] from  -.58 .81  to -1.3 1.5 
\arrow <.35cm> [.1,.4] from  0 0 to  -.087 1.0 
\arrow <.35cm> [.1,.4] from  -.087 1.0 to .0013 2.0
\arrow <.35cm> [.1,.4] from  0 0 to  0.0 1.0  
\arrow <.35cm> [.1,.4] from  0.0 1.0 to .257 1.97 
\arrow <.35cm> [.1,.4] from  0 0 to  1.0 0.0  
\arrow <.35cm> [.1,.4] from  1.0 0.0 to 1.99 -0.13
\arrow <.35cm> [.1,.4] from  0 0 to  -1.0 0.0  
\arrow <.35cm> [.1,.4] from  -1.0 0.0 to -1.99 0.13

\put {\mathput{n}}                at  .28 2.125 
\put {\mathput{\phi}}             at  2.15 -0.13 
\put {\mathput{-\phi}}            at  -2.2 0.13 
\put {\mathput{\overline{\rm car}}} at  2.20 0.05
\put {car}                        at  .65 2.05  
\put {\mathput{\mathcal{Z}+}}          at  1.2 1.8 
\put {\mathput{\mathcal{Z}-}}          at  -1.3 1.7 
\put {geo$-$}                     at  -1.80 1.45 
\put {geo$+$}                     at  1.8 1.4 
\put {null$+$}                     at 2.05  1.05  
\put {null$-$}                     at -1.85 1.25   
\put {\mathput{t}}                at  .06 2.17 
\put {gmp}                        at  -.2 2.15 
\put {\bf ext}                        at  -.65 2.1 
\put {\mathput{\overline{\bf ext}}} at  -2.2  .55

\setquadratic
\setdashes
\plot 1.99 -0.13
2. 0. 1.99 .0742 1.97 .251 1.92 .448 1.86 .643 1.77 .835 1.66 1.01 1.60 1.10 1.53 1.19 1.45 1.28 1.36 1.37 1.27 1.46 1.16 1.55 1.04 1.64 .902 1.73 .769 1.80 .624 1.86 .446 1.92 .256 1.97 .0556 1.99 -.148 1.99 -.340 1.97 -.524 1.93 -.681 1.88 -.827 1.82 -.984 1.74 -1.12 1.65 -1.23 1.57 -1.33 1.49 -1.43 1.39 -1.52 1.29 -1.59 1.20 -1.66 1.11 -1.72 1.01 -1.78 .912 -1.82 .817 -1.86 .721 -1.90 .621 -1.93 .520 -1.95 .428 -1.97 .335 
-1.99 .131 
 /
\endpicture}$$

\caption{The orientation of the intersection of the parallel transport Lorentz transformation plane with the $t$-$\phi$ tangent plane for the various directions inside the half circle for a Kerr black hole with 
$a/\mathcal{M}=.5$ at $r/\mathcal{M}=4$. 
The horizontal and vertical directions are along $e_{\hat\phi}$ and $n$ (ZAMO angular direction and 4-velocity) respectively, while the null directions (dashed lines) are at 45 degrees from the horizontal. Each arrow inside the half circle indicates a direction in the $t$-$\phi$ velocity space and the direction arrow attached to it indicates the direction given by the composed map $\bar\mathcal{Z}$ applied to the corresponding relative velocity.  The direction of the covariant constant vectors for a given initial direction is the direction orthogonal to the attached arrow in the Lorentz geometry (reflection across the 45 degree line).
The double arc indicates the boost zone. The extremely accelerated observer direction and its orthogonal counterpart indicated by {\bf ext} and $\overline{\bf ext}$ are the only ones aligned with their boost/rotation plane direction, while the two geodesics are the only ones aligned with their covariant constant vector direction. The maps $\mathcal{Z}$ and $\bar\mathcal{Z}$ are actually both reflection symmetric across the extremely accelerated observer direction.}
\label{fig:1}
\end{figure}

The exponentiation for the solution of the parallel transport equations 
\beq
X(\phi) =e^{\phi A(\zeta)} X(0)
\eeq
and its interpretation as a Lorentz transformation in a 2-plane is again accomplished with the relations
\begin{eqnarray}\label{eq:ABzeta}
\fl\qquad&&
A(\zeta) = \sigma(\zeta) B(\zeta)
         =  \sigma_{(0)}(\zeta) \mathcal{B}(\zeta)
  \ ,\nonumber\\
\fl\qquad&&
\sigma(\zeta) = \gamma(\mathcal{Z}(\zeta))^{-1} [R/\rho]\,                      
         (1-\zeta_{\rm(gmp)}/\zeta) \ ,\qquad
\sigma_{(0)}(\zeta) =  [R/\rho]\,                      
         (1-\zeta_{\rm(gmp)}/\zeta)
\ ,\nonumber\\
\fl\qquad&&
B(\zeta)^\flat 
  = e_{\hat r}^\flat \wedge \bar U(\mathcal{Z}(\zeta))^\flat \ ,\qquad 
\mathcal{B}(\zeta)^\flat 
  = e_{\hat r}^\flat \wedge Y(\mathcal{Z}(\zeta))^\flat
\ ,\nonumber\\
\fl\qquad&&
B(\zeta)^3 = -\epsilon(\mathcal{Z}(\zeta))^2 B(\zeta)\ ,\qquad
\mathcal{B}^3= -\epsilon(\mathcal{Z}(\zeta))^2  \gamma(\mathcal{Z}(\zeta))^{-2} \mathcal{B}(\zeta) \ .
\end{eqnarray}

The general exponential solution then has exactly the same form as (\ref{eq:expB}), in terms of the $\zeta$-generalized quantities 
\beq\label{eq:expBzeta}
\fl
 e^{\phi A(\zeta)}
 = \cases{ 
e^{\phi\sigma(\zeta) B(\zeta)}
= [1-\mathcal{P}(\zeta)]
  + \cos[\epsilon(\mathcal{Z}(\zeta))\sigma(\zeta)\phi] 
  \, \mathcal{P}(\zeta) &\cr
  \null\hskip50pt + \epsilon(\mathcal{Z}(\zeta))^{-1} 
    \sin[\epsilon(\mathcal{Z}(\zeta))\sigma(\zeta)\phi] 
  \, B(\zeta) \,, 
   & \kern-20pt
\hbox{$\epsilon(\mathcal{Z}(\zeta))\neq0$\,,}\cr
e^{\phi\sigma_{(0)}(\zeta) \mathcal{B}(\zeta)}
= 1 + \sigma_{(0)}(\zeta)\phi \mathcal{B}(\zeta) + \frac12\sigma_{(0)}(\zeta)^2\phi^2 \mathcal{B}(\zeta)^2\,, 
   & \kern-20pt\hbox{$\epsilon(\mathcal{Z}(\zeta))=0$\,,}\cr
}
\eeq
where now
\beq
 \mathcal{P}(\zeta) = -\epsilon(\mathcal{Z}(\zeta))^{-2} B(\zeta)^2 
\ ,\qquad \epsilon(\mathcal{Z}(\zeta))\neq0
\ .
\eeq

Note that by defining the new dimensionless frequency with respect to the GMPOs by
$\tilde\sigma(\zeta) \tilde\phi= \sigma(\zeta)\phi$, one finds
\begin{eqnarray}
\fl\qquad
   \tilde\sigma(\zeta) 
     &=& (1-\zeta_{\rm(gmp)}/\zeta)^{-1} \sigma(\zeta)
     = \gamma(\mathcal{Z}(\zeta))^{-1} R/\rho \ ,
  \qquad \epsilon(\mathcal{Z}(\zeta)) \neq0\ ,
\nonumber\\
\fl\qquad
   \tilde\sigma_{(0)}(\zeta) 
     &=& (1-\zeta_{\rm(gmp)}/\zeta)^{-1} \sigma_{(0)}(\zeta) 
     = R/\rho \ ,
  \qquad \kern35pt\hbox{$\epsilon(\mathcal{Z}(\zeta)) =0$} \ ,
\end{eqnarray}
so it is the angle with respect to the GMPOs which plays the same role as the coordinate angle for the $\phi$-coordinate circle discussion. When $\mathcal{Z}(\zeta)$ corresponds to a timelike direction, the inverse gamma factor in $\tilde\sigma$ locally Lorentz contracts spacetime arclength along the $\phi$-coordinate circle to spacetime arclength along the corresponding tilted local rest space direction.

The local direction of the plane of the Lorentz transformation within the $t$-$\phi$ cylinder is now along $\bar U(\mathcal{Z}(\zeta))$ (when $\epsilon(\mathcal{Z}(\zeta))\neq0$) or $Y(\mathcal{Z}(\zeta))$ (when $\epsilon(\mathcal{Z}(\zeta))=0$). 
For the timelike circular geodesics, which are invariant under the map $\mathcal{Z}$, 
this is just the angular direction in their local rest space, while along the time lines this is the local time direction of the Carter observers. In general when the $\mathcal{Z}(\zeta)$ trajectories are timelike/spacelike, this represents a rotation/boost; the horizon for the corresponding observers is reached when  $\gamma(\mathcal{Z}(\zeta))^{-1}$ vanishes. This occurs at the two angular velocities $\zeta=\mathcal{Z}(\zeta_\pm)$ satisfying the inequalities (\ref{eq:Zzeta}). At a given radius, the interior of this angular velocity interval represents the boost zone, while its exterior represents the rotation zone, separated by null rotations which occur at the endpoints. 

Recall that $\partial_t+\mathcal{Z}(\zeta)\partial_\phi$ is the direction of the stationary axisymmetric vectors which are tangent to the $t$-$\phi$ cylinder and invariant under this Lorentz transformation and hence are covariant constant along the curve. 
The extremely accelerated observer world lines are the only circular orbits for which this covariant constant angular direction is orthogonal to the orbits themselves, corresponding to the phase-locking of Fermi-Walker transported spin vectors found by de Felice and Usseglio-Tomasset \cite{defuss}.
In fact since $\mathcal{Z}$ must be a fractional linear transformation of the relative velocity with respect to any observer (since the velocity addition boost formula is a fractional linear transformation),
the four conditions on the first line of (\ref{eq:Zconditions}) show that with respect to the extremely accelerated observers
$\mathcal{Z}: \nu \to -\nu_-\nu_+/\nu$ (where $\nu_- = -\nu_+$). Thus these observers see both the geodesic relative velocities and the map $\mathcal{Z}$ symmetrically in the relative speed $|\nu|$. 

If one considers circular orbits with a constant (finite) angular velocity, the orbits are not closed and one cannot consider spacetime holonomy. One can obtain a closed loop by joining together two distinct orbits of this type at a crossing point, which then periodically meet each other, forming a sequence of closed loops in spacetime. For example, in the Kerr spacetime one could choose the first orbit to have an arbitrary angular velocity $\zeta$ and the second orbit to be a time coordinate line with zero angular velocity, which is a world line of a static observer, along which vectors are boosted by parallel transport. The net holonomy transformation would then be the inverse of one Lorentz transformation times the other Lorentz transformation.
Rothman et al \cite{rot}
slide over this point by moving quickly on to the case in which the two orbits are oppositely-rotating geodesics.

It should be noted that Marck has explicitly constructed a parallel transported frame for general geodesics in the Kerr black hole using its Killing-Yano tensor, a result which extends to a wider class of stationary axisymmetric spacetimes \cite{marck}.

\section{Timelike geodesic holonomy}

Along the timelike geodesics starting at $t=0$ and $\phi=0$ and parametrized by $t$, 
the parametrizations by the GMPO comoving angle and the coordinate angle
and the corresponding dimensionless frequencies are now related to each other by 
\beq
  \frac{\tilde\phi_\pm}{\phi_\pm}
       = (1- \zeta_{\rm(gmp)}/\zeta_{\rm(geo)\pm})
    = (1 \mp a\omega_{(0)} )^{-1} 
=\frac{\sigma(\zeta_{\rm(geo)\pm})}{\tilde\sigma(\zeta_{\rm(geo)\pm})}
\ ,
\eeq
while 
\beq
\tilde\sigma(\zeta_{\rm(geo)\pm}) =  \gamma_\pm^{-1} R/\rho\ .
\eeq
It is convenient to introduce the abbreviations 
\beq\label{eq:sigmaf}
\fl
\sigma_\pm=\sigma(\zeta_{\rm(geo)\pm})\ ,\
\sigma_{\rm(avg)} = (\sigma_+ + \sigma_-)/2\ ,\
\sigma_{\rm(dif)} = (\sigma_+ - \sigma_-)\ ,\
f_\pm^2=\epsilon(\zeta_{\rm(geo)\pm})^2 \sigma_\pm^2
\eeq
and their corresponding tilde versions.
The quantities $f_\pm^2$ 
are the frequency-squared functions appearing in the decoupled second-order equations analogous to (\ref{decoupled}) for parallel transport along the circular geodesics.
For the Kerr spacetime because of the special condition (\ref{eq:Kerr-condition}),
one has
\begin{eqnarray}
\sigma_\pm 
&=&  |1-\frac{3\mathcal{M}}{r} \pm 2a\omega_{(0)} |^{1/2}
 =  \frac{\tau_\pm}{T_{(0)}}
\ ,\nonumber\\
\tilde\sigma_\pm
&=& |1-\frac{3\mathcal{M}}{r} \pm 2a\omega_{(0)}|^{1/2}  (1\mp a\omega_{(0)})
= \frac{\tilde\tau_\pm}{T_{(0)}}
\ .
\end{eqnarray}
Thus in the Kerr spacetime
the dimensionless frequency factors which convert the angle of revolution with respect to the GMPOs to the parallel transport angle with respect to the boosted spherical frame are just the ratios of the individual geodesic proper periods to the average coordinate period with respect to the static observers.
The corrected frequency factor $\sigma_\pm -1$ corresponds to the rotational angle with respect to boosted spherical axes whose orbital rotation is removed \cite{idcf2,vis,rin}.

These frequency functions ($\tilde\sigma_\pm$ and $\sigma_\pm$) go to zero when the corresponding geodesics become null as $\gamma_\pm^{-1}\to 0$, 
corresponding to the single GMP horizon in the case of the corresponding discussion for the $\phi$-coordinate circles.
For the Kerr spacetime, these `geodesic horizons'
mark the regions in the equatorial plane where as one approaches the black hole, first the counterrotating geodesic goes null 
(at $r=r_{AB}$, where $\sigma_-=0$) 
and then the corotating geodesic goes null, 
(at $r=r_{BC}$, where $\sigma_+=0$), 
which divides the equatorial plane into three regions A, B, C, where of the two circular geodesics respectively both are timelike or one or neither is timelike. Each such radius represents the only zero of the corresponding frequency function outside the horizon.
Explicit formulas for these two radii are given by Chandrasekhar \cite{chandra}. 
Figure 4 of \cite{idcf2} shows a plot of these radii and the horizon and ergosphere radii as a function of $a$ for the physical interval $a\in[0,1]$. Figure 2 reproduces these curves together with the GMPO horizon radius $r_{\rm(gmp)}=r_*$. 

\typeout{MAPLE figure}

\begin{figure}
\typeout{*** EPS figure 1}
\centerline{
\epsfbox{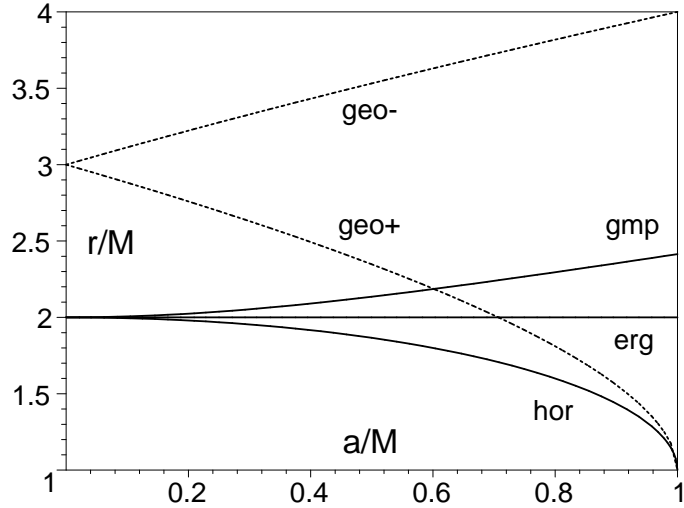}
}
\caption{Radius ratio $r/\mathcal{M}$ versus $a/\mathcal{M}$ in the Kerr equatorial plane for the horizon radii of the ZAMOs (hor), the static observers (erg), the GMPOs (gmp), and the counterrotating and corotating geodesic observers (geo$-$, geo$+$). 
As $a/\mathcal{M}$ increases from 0 to 1, the horizon radius ratio $r/\mathcal{M}$ decreases from 2 to 1, the ergosphere ratio remains at 2, and the GMP horizon ratio increases from 2 to $1+\sqrt{2}$. The curves marked `geo$-$' and `geo$+$' are the radii of the counterrotating and corotating null circular geodesics, which start at 3 and end at 4 and 1 respectively.}
\label{fig:2}
\end{figure}

When both circular geodesics are timelike,
the parallel transport Lorentz transformation is a rotation by the respective angles $\tilde\sigma_\pm \tilde\phi_\pm$, but this rotation takes place in different planes, namely in the local rest space of each geodesic.
The geodesics meet at every half-revolution with respect to the GMPOs; the net parallel transport angles along each are
\beq
  \pm  \pi \tilde\sigma_\pm 
= \pm  \pi \gamma_\pm^{-1} [R/\rho] \ .
\eeq
Apart from the gamma factor, this corresponds exactly to the shortened rotational angle by which a transported vector rotates in the flat tangent cone to the embedding surface during each half revolution with respect to the ZAMOs, and the gamma factor then Lorentz contracts this amount to correspond to the angle in the geodesic local rest space which projects down onto this angle in the ZAMO local rest space.

After a positive integer $q$ of such half-revolutions,
at either the original GMPO world line ($q$ even) or the antipodal GMPO world line ($q$ odd),
the net parallel transport angles along each are
\beq
  \pm q \pi \tilde\sigma_\pm 
= \pm q \pi \gamma_\pm^{-1} [R/\rho] \ .
\eeq
Joining two oppositely-rotating geodesic curves together at their common departure point (initial meeting point), one obtains a closed loop representing $q$ complete circuits with respect to the GMPOs. With a common initial vector $X(0)$ at this starting point at $\phi=0=\tilde\phi$, the two parallel transported vectors $X(\phi_\pm)$ at the final meeting point will differ by a holonomy transformation belonging to the Lorentz group.

The orthogonal projections of $X(\phi)$ into the two local rest spaces of the pair of oppositely-rotating geodesics define its apparent directions as seen by those geodesics. Only this orthogonal component evolves along each geodesic, undergoing a rotation in a fixed (coordinate component) plane. The holonomy transformation at the final meeting point then results from applying the inverse of one of these rotations, followed by the boost from that local rest space to the other, and then applying the other rotation. At those radii for which this transformation is the identity matrix, one finds holonomy invariance.
Holonomy invariance for the entire tangent space requires that each rotation separately return to the identity, since the two rotations take place in different planes 
and the only way one can reassemble the two pairs of orthogonal pieces of the final vectors $X(\phi_\pm)$ after parallel transport to form the same spacetime vector if the temporal parts are unchanged is if the spatial parts are also unchanged.

In the static case, where $\zeta_{\rm(gmp)}=0$ and $\tilde\sigma_\pm=\sigma_\pm$,  
the geodesic gamma factors $\gamma_+=\gamma_-$
are the same and the two rotations in opposite senses by the same angle in absolute value will agree if they are each equal to a rotation by some multiple of $2\pi$, so
\beq
 \gamma_\pm^{-1}  [R/\rho]\, q\pi
 = 2\pi p 
\quad \hbox{or}\quad
 \tilde\sigma_\pm
     =  
 \gamma_\pm^{-1}  R/\rho 
  = \frac{2p}{q}\ ,
\eeq
where $p$ is some positive integer representing the number of parallel transport loops which take the the initial vector to the final vector along each such  
curve separately (with respect to the boosted spherical frame along the way).
This occurs when the GMP frequency functions 
$\tilde\sigma_\pm$
assume rational values of the form $2p/q$.
If $R/ \rho <1$ as in the static Kerr case (Schwarzschild), one must have 
$2p<q$ and $q>2$. 
Here the formula (\ref{eq:gammaS}) reduces
to $\gamma_\pm = (1-2\mathcal{M}/r)^{1/2}/(1-3\mathcal{M}/r)^{1/2}$, thus 
converting the 
$\gamma(\zeta_{\rm(gmp})^{-1} R/\rho 
= R/\rho  =(1-2M/r)^{1/2}$ 
factor for $\phi$-coordinate orbits to 
$\gamma_\pm^{-1}  R/\rho = (1-3M/r)^{1/2}$ for geodesics, recovering both results of Rothman et al \cite{rot} for the frequency functions, but they neglect the crucial factor of 2 which rules out the odd values of their parameter $m = 2p$. 
Note that $2p$ is the total number of loops with respect to the boosted spherical frame that the final vector $X(\phi_-)$ undergoes as it is parallel transported back from the final meeting point to $X(0)$ along $q$  counterrotating half-circuits
and then on to the final vector $X(\phi_+)$ after $q$ corotating half-circuits returning to the final meeting point.
In this way the `even' members of the harmonic parametrization $q=1,2,\ldots$, $p=1,2,\ldots,q-1$ generate a corresponding discrete spectrum of radii at which the holonomy subgroups restricted to these orbits are finite.
These radii densely cover the radial interval where the two circular geodesics are timelike, reproducing the band structure of holonomy invariance found for the closed $\phi$-coordinate loops.

These geodesic gamma factors are distinct in the nonstatic case 
($\sigma_+\neq\sigma_-$),
where the condition that the separate rotations each return to the identity leads to two independent conditions rather than one as in the static case
\beq
 \tilde\sigma_\pm \, q\pi
 = 2\pi p_\pm 
\quad \to \quad
 \tilde\sigma_\pm =
 2p_\pm /q\ ,
\eeq
where now $p_+\neq p_-$.
In general it is not possible to satisfy both conditions at the same radius, although exceptional solutions of this type can occur at a set of radii of measure zero in the interval where both geodesics are timelike.
This is clearly seen in the Kerr spacetime case (see Fig.~2)
where as one increases the rotation parameter $a$ away from 0, the graphs of $\tilde\sigma_\pm$ versus $r$ move away on opposite sides from the Schwarzschild graph. The counterrotating geodesic frequency curve moves to the right like the corresponding horizon $r_{AB}$ of the counterrotating geodesics,  while the cororating geodesic frequency curve moves to the left like the corresponding horizon $r_{BC}$ of the corotating geodesics, with their zero values marking these two respective radii. 
For fixed $q$ and a given pair $p_+<p_-$ in Schwarzschild, the corresponding radii $r_{(q,p_\pm)}$ at which $\tilde\sigma_\pm=2p_\pm/q$
satisfy  $r_{(q,p_+)}< r_{(q,p_-)}$, but as one increases the rotation parameter $a$ for a given $q$, the various radii pairs eventually cross, one by one, modulo large $a$ effects. However, for a given value of $a$, there is no guarantee that even one such common radius exists.  
Thus the band structure of holonomy invariance which exists for the static case does not exist in the nonstatic case, where it has instead been diluted by spreading out over the parameter interval of $a$ values, unlike the case $p_+=p_-$, which all occur at the single value $a=0$. A similar situation probably arises for other spacetimes of this clock-effect family.

Figure 3 illustrates this situation for the Kerr frequency functions with $a/\mathcal{M}=0.5$ compared to the Schwarzschild case (middle curve). The horizontal lines at $\frac14,\frac24,\frac34$ illustrate either a $q=4$ circuit loop and $2p_\pm=1,2,3$ or a $q=8$ circuit loop and $2p_\pm=2,4,6$.
For example, the intersection points with the lines at $2p_\pm/8$ mark the radii at which the parallel transport rotation separately returns to the identity compared to the spherical frame after an 8-circuit loop, in the process making $p_\pm=1,2,3$ revolutions with respect to that frame. One can see that the pair $(p_+,p_-)=(2,1)$ has already crossed at a smaller value of $a$, while the pairs $(3,1)$ and $(3,2)$ have yet to cross for larger values of $a$, all of which are easily found numerically.

\typeout{MAPLE figure}

\begin{figure}
\typeout{*** EPS figure 2}
\centerline{
\epsfbox{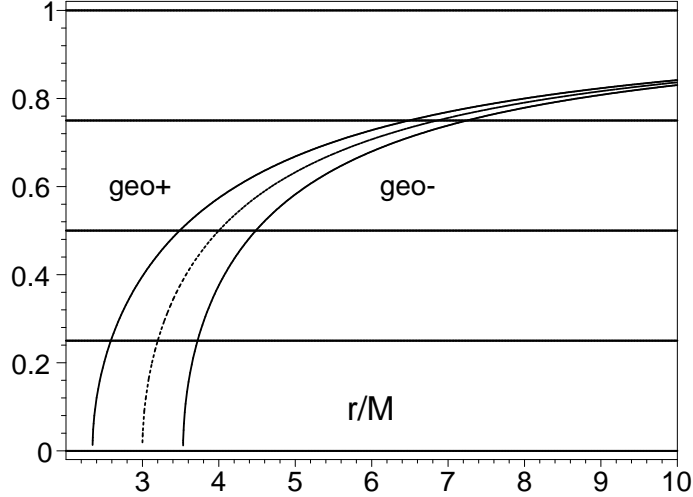}
}
\caption{The GMP frequency functions $\tilde\sigma_\pm$
versus $r/\mathcal{M}$ in the Kerr equatorial plane for the corotating and counterrotating geodesic observers (geo$+$, geo$-$) in the 
the Kerr spacetime with $a/\mathcal{M}=0.5$, together with the  $a=0$ limit (middle curve). 
The black hole horizon is at $r/\mathcal{M}=2$.
Intersections with the three equally spaced horizontal lines between 0 and 1 mark the radii at which the parallel transport rotations separately return to the identity after $q=8$ GMP circuits corresponding to  $p=1,2,3$ (left to right) complete revolutions with respect to the spherical frame.
}
\label{fig:3}
\end{figure}

The real problem here is the boost between the local rest spaces of the pair of geodesics. If one were interested in comparing Fermi-Walker transport along curves modulo such boosts, then an analog of holonomy invariance still exists. Fermi-Walker transport along a timelike curve describes the behavior of the spin of a test gyroscope carried along such a world line, and reduces to parallel transport along a geodesic. Spin vectors lie in the local rest space of the world line along which they travel, so to compare, at a spacetime point, spin vectors carried by observers in relative motion, one has to boost the spin vector from one observer to the other. 

The degree to which this is physically interesting can be disputed, but if one accepts it mathematically, then one can parallel transport a spin vector around a closed geodesic loop like a baton in a relay race, passing off the baton from one geodesic to the other at the start and finish by a boost in order to make the comparison in the orientation of the spin direction without the relative motion distortion. For the case at hand, the boost comparison identifies the angular direction along $\phi$ in the local rest spaces (namely $\bar U(\zeta_\pm)$), allowing one to ignore the fact that the rotations take place in different planes. 

On the other hand one can imagine an experimental situation where each spin system is interrogated separately at the crossing event.
Then the final rotations will agree (ignoring the boost) if the two separate angles are the same mod $2\pi$ after $q>0$ separate half-circuits (joint whole circuits)
\beq
 q\pi \, \tilde\sigma_{\rm(geo)+}-(-q\pi \, \tilde\sigma_{\rm(geo)-}  )  
 = 2 q\pi \, \tilde\sigma_{\rm(avg)}  
 = 2\pi p
\ \to\
\tilde\sigma_{\rm(avg)} = \frac{p}{q}\ ,
\eeq
i.e., when the average frequency function $\tilde\sigma_{\rm(avg)}$
assumes rational values. This leads to a situation analogous to the static case, but now including the odd values of the numerator $p$.
The average of the two individual angles instead gives the relative angle of their common direction (modulo the boost) with respect to the spherical frame at the meeting point
\beq
 \frac12[q\pi \, \tilde\sigma_{\rm(geo)+}+(-q\pi \, \tilde\sigma_{\rm(geo)-}  )] 
= \frac12 q\pi\,\tilde\sigma_{\rm(dif)}  
= \frac12 q\pi \,\sigma
    \frac{\tilde\tau_+ -\tilde\tau_-}{\tilde\tau_{(0)}}\ ,
\eeq
which is directly proportional to the fractional two-clock clock effect given by the ratio of (\ref{eq:2c}) and (\ref{eq:2period}). The proportionality factor is just the frequency function $\sigma$
for the $\phi$-coordinate circle discussion multiplied by $\frac12 q\pi$.

If one imagines a gedanken dual drag-free satellite gyroscope experiment with two oppositely-rotating equatorial plane geodesic orbits at the same radius, then it makes sense to assume that the internal spin up mechanisms would give the gyros the same spin magnitude in their rest frame relative to the satellite housing itself, which is presumably locked onto a distant guide star and hence to the local spherical frame in a calculable way. Since each spin vector on these orbits precesses around an axis fixed with respect to the boosted spherical frame,
one could compare the total angles of precession at meeting points with each other as well as with the local spherical frame, in which case the above mathematical considerations would come into play, including the use of boosts to compare orientations of gyros and axes in relative motion. These individual rotations of the spin vectors represent the combined 
geodetic and Lense-Thirring precession effects, which are aligned for these orbits, in contrast with slowly precessing polar orbits in the limit of small $a$ where they are orthogonal, as in the GP-B experiment \cite{gpb}.

\section{Conclusions}

Previous work on the holonomy associated with circular orbits in particular stationary axisymmetric spacetimes has been generalized to all those which admit clock effects, using the Kerr spacetime as an illustrative example. The details of the calculations depend heavily on the geometry of the GMPOs which are themselves defined by the geometry of clock effects. They directly determine the parallel transport transformations relative to the natural symmetry-adapted frame along the closed curves tangent to the Killing vector associated with axial symmetry, and do the same in combination with the Carter observers for general circular orbits.
In the nonstatic case the band of holonomy invariance which exists for special geodesic clock-effect loops in the static case disappears.
With some modifications, one can extend these parallel transport results to the case of spin precession to compare the total precession angles.
In the Kerr spacetime the dimensionless rotational frequencies associated with parallel transport rotations are particularly simple ratios of various circular geodesic period quantities.
The rich mathematical structure that has emerged from this geometric problem has been made apparent only through the relative observer analysis associated with gravitoelectromagnetism. 

\section*{Acknowledgements}

D.B. acknowledges fruitful discussions with  Dr.~G. Ali at the Istituto per Applicazioni della Matematica.

\appendix

\section{Frenet-Serret approach}

Accelerated curves have a natural orthonormal frame, the Frenet-Serret frame, with respect to which the parallel transport Lorentz transformation is described in terms of the curvature $\kappa$ (magnitude of acceleration) and the first and second torsions $\tau_1$ and $\tau_2$ (Fermi-Walker rotation parameters). The relation between this analysis and the coordinate frame analysis performed here is rather simple.

The general Frenet-Serret frame $E_\alpha$
along a circular orbit \cite{vish1,vis,vish3,circfs}, for which $E_0=U(\zeta)$, satisfies
\beq\fl\qquad
 \frac{D E_\alpha}{d\tau} = F^\beta{}_\alpha E_\beta\ ,\ 
  F^{\alpha\beta} = [\kappa E_0\wedge E_1 -\tau_1 E_1\wedge E_2 
                         -\tau_2 E_2\wedge E_3]^{\alpha\beta}\ ,
\eeq
which is explicitly
\beq\fl 
   \frac{DE_0}{d\tau} = \kappa E_1\ ,\ 
   \frac{DE_1}{d\tau} = \kappa E_0 +\tau_1 E_2\ ,\ 
   \frac{DE_2}{d\tau} = -\tau_1 E_1 +\tau_2 E_3\ ,\ 
   \frac{DE_3}{d\tau} = -\tau_2 E_2\ .
\eeq
A vector $Z=Z^\alpha E_\alpha$ that is  parallel transported along $U$ then satisfies
\beq
\frac{D Z^\alpha}{d\tau} = \frac{d Z^\alpha}{d\tau} + F^\alpha{}_\beta Z^\beta\ .
\eeq 
For this case the frame is given by
$
  E_0 = U(\zeta)\,,\,
  E_1 = e_{\hat r}\,,\,
  E_2 = \bar U(\zeta)\,,\,
  E_3 = e_{\hat \theta}\,,\,
$
and the curvature and torsions are
\beq
\fl\qquad
\kappa = -\rho^{-1} \gamma(\zeta)^2 (\nu(\zeta)-\nu_+) (\nu(\zeta)-\nu_-)\ ,\
\tau_1 = -\frac12 \gamma(\zeta)^{-2}\frac{d\kappa}{d\nu}\ ,\
\tau_2=0\ ,
\eeq
leading to the Frenet-Serret relations
\begin{eqnarray}
\fl\qquad
   \frac{D U(\zeta)}{d\tau} = \kappa e_{\hat r}\ ,\ 
   \frac{D e_{\hat r}}{d\tau} = \kappa U(\zeta) +\tau_1 \bar U(\zeta)\ ,
\ 
   \frac{D\bar U(\zeta)}{d\tau} = -\tau_1 e_{\hat r} \ ,\ 
   \frac{D e_{\hat \theta}}{d\tau} = 0\ .
\end{eqnarray}

From the definitions (\ref{eq:Azeta}) and (\ref{eq:ABzeta}), which are  equivalent to 
\beq\label{eq:Der}
  \frac{D e_{\hat r}}{d\phi} = \sigma(\zeta)\bar U(\mathcal{Z}(\zeta))\ ,
\eeq
and  the second Frenet-Serret relation together with the reparametrization relation $d\phi/d\tau=\gamma(\zeta) \zeta/N$, one has, when $U(\zeta)$ and $U(\mathcal{Z}(\zeta))$ are nonnull,
\begin{eqnarray}\label{eq:ktU}
\fl\qquad
\bar U(\mathcal{Z}(\zeta)) 
&=& \pm |\tau_1^2 - \kappa^2|^{-1/2} [\kappa U(\zeta) + \tau_1 \bar U(\zeta)]
\ ,\nonumber\\
\fl\qquad
     U(\mathcal{Z}(\zeta)) 
&=& \pm |\tau_1^2 - \kappa^2|^{-1/2} [\tau_1 U(\zeta) + \kappa \bar U(\zeta)]
\ ,\nonumber\\
\fl\qquad
\sigma(\zeta) \gamma(\zeta)\zeta/N 
&=& \pm |\tau_1^2 - \kappa^2|^{1/2}
      = \pm |F^{\alpha\beta} F_{\alpha\beta}/2|^{1/2}
\ ,
\end{eqnarray}
modulo signs which depend on the signs of $\kappa,\tau_1,\zeta$ and where
\beq\fl\qquad
|\tau_1| =\frac{1}{\rho}
[(\nu -\nu_{\mathcal{Z}(\zeta_+)}) (\nu -\nu_{\mathcal{Z}(\zeta_-)})]^{1/2}
\frac{\gamma (\zeta) \, \gamma (U(\mathcal{Z}(\zeta)),U(\zeta))}
         {\gamma_{\rm (gmp)}}
\ .
\eeq
Modulo these signs the first two lines of (\ref{eq:ktU}) represent a boost  from $(U(\zeta),\bar U(\zeta))$ to $(U(\mathcal{Z}(\zeta)),\bar U(\mathcal{Z}(\zeta)))$ with relative velocity
$\kappa/\tau_1$ when $|\kappa/\tau_1|<1$, and a boost plus the bar map when  $|\kappa/\tau_1|>1$.

The case of  Fermi-Walker transport instead of parallel transport corresponds to the replacements
$D\to D_{\rm(fw)}$, $\kappa\to0$ in the above relations,
while projection of (\ref{eq:Der}) into the local rest space of 
$U(\zeta)$ 
yields the Fermi-Walker derivative along $U(\zeta)$ of the vector field $e_{\hat r}$, which is spatial with respect to all circularly rotating observers
\begin{eqnarray}\label{eq:PDer}
\fl
  \frac{D_{\rm(fw)} e_{\hat r}}{d\phi}  
  = P(U(\zeta))\frac{D e_{\hat r}}{d\phi}
= \sigma(\zeta)P(U(\zeta))\bar U(\mathcal{Z}(\zeta))
= \sigma(\zeta)\gamma(U(\mathcal{Z}(\zeta)),U(\zeta)) \, \bar U(\zeta)\ .
\end{eqnarray}
Thus the Fermi-Walker transport dimensionless frequency is instead
\beq
  \sigma_{\rm(fw)}(\zeta)
   = \gamma(U(\mathcal{Z}(\zeta)),U(\zeta)) \,\sigma(\zeta)\ ,
\eeq
with an extra relative gamma factor describing the inverse Lorentz contraction which occurs in the projection, which has unit value for geodesics where this projection reduces to the identity.
The relative gamma factor may be expressed in terms of the ZAMO gamma factors \cite{rok} as 
$
\gamma(U(\mathcal{Z}(\zeta)),U(\zeta))
= \gamma(\mathcal{Z}(\zeta)) \gamma(\zeta)
  [1-\nu(\mathcal{Z}(\zeta)) \nu(\zeta)]
$.


\end{document}